\documentstyle[12pt,aps,epsfig]{revtex}

\parskip=\medskipamount

\def\be{\begin{equation}}
\def\ee{\end{equation}}
\def\disp{\displaystyle}

\def\R{{\sf I\kern-.15em R}}
\def\T{{\sf T\kern-.45em T}}
\def\C{\kern.1em{\raise.47ex\hbox{$\scriptscriptstyle |$}}
             \kern-.40em{\sf C}}
\def\Z{{\sf Z\kern-.45em Z}}
\def\vtheta{\vartheta}

\def\hfl#1#2{\smash{\mathop{\hbox to
10mm{\rightarrowfill}}\limits^{\scriptstyle#1}_{\scriptstyle#2}}}

\begin{document}

\begin{titlepage}

\title
{\Large \bf On the Plants Leaves Boundary, ``Jupe \`a Godets'' \\ and
Conformal Embeddings}

\author
{Sergei Nechaev$^{\dag\,\ddag\,1}$, Rapha\"el Voituriez$^{\dag\,2}$}

\address
{\it $^{\dag}$ Laboratoire de Physique Th\'eorique et Mod\`eles
Statistiques, Universit\'e Paris Sud, \\ 91405 Orsay Cedex, France \\
$^{\ddag}$ L D Landau Institute for Theoretical Physics, 117940, Moscow, 
Russia}


\maketitle

\begin{abstract}

The stable profile of the boundary of a plant's leaf fluctuating in the
direction transversal to the leaf's surface is described in the framework
of a model called a "surface \`a godets". It is shown that the information
on the profile is encoded in the Jacobian of a conformal mapping (the
coefficient of deformation) corresponding to an isometric embedding of a
uniform Cayley tree into the 3D Euclidean space. The geometric
characteristics of the leaf's boundary (like the perimeter and the height)
are calculated. In addition a symbolic language allowing to investigate
statistical properties of a "surface \`a godets" with annealed random
defects of curvature of density $q$ is developed. It is found that at $q=1$
the surface exhibits a phase transition with the critical exponent
$\alpha=\frac{1}{2}$ from the exponentially growing to the flat structure.

\end{abstract}

\vspace{0.5in}

\noindent
{\bf key words:} isometric embedding, tellesation, conformal transform, 
graph of the group

\bigskip


\vspace{3in}

\hrule
\bigskip

\noindent
$^1$ E-mail: nechaev@ipno.in2p3.fr \\ 
$^2$ E-mail: voiturie@ipno.in2p3.fr

\end{titlepage}

\section{Introduction}
\label{sect:1}

The very subject of this paper has been inspired by the following
questions raised by V.E.Za\-kha\-rov in a private conversation about two
years ago: (1) "What are the geometrical reasons for the boundary of plants
leaves to fluctuate in the direction transversal to the leaf's surface?"
and (2) "How to describe the corresponding stable profile?" The answer to
question (1) has came immediately.

\begin{figure}[ht]
\begin{center}
\epsfig{file=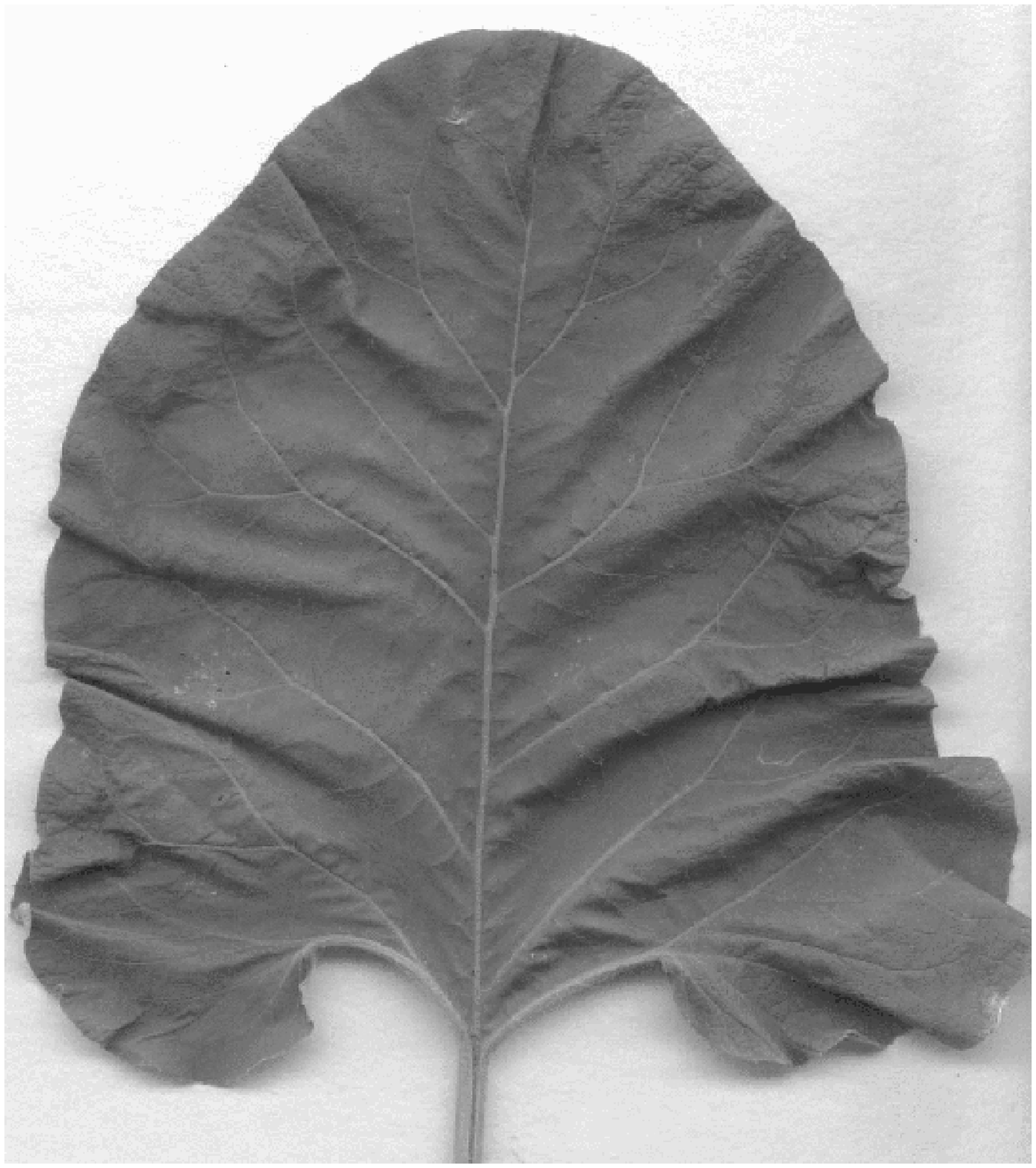,width=6cm,angle=0}
\epsfig{file=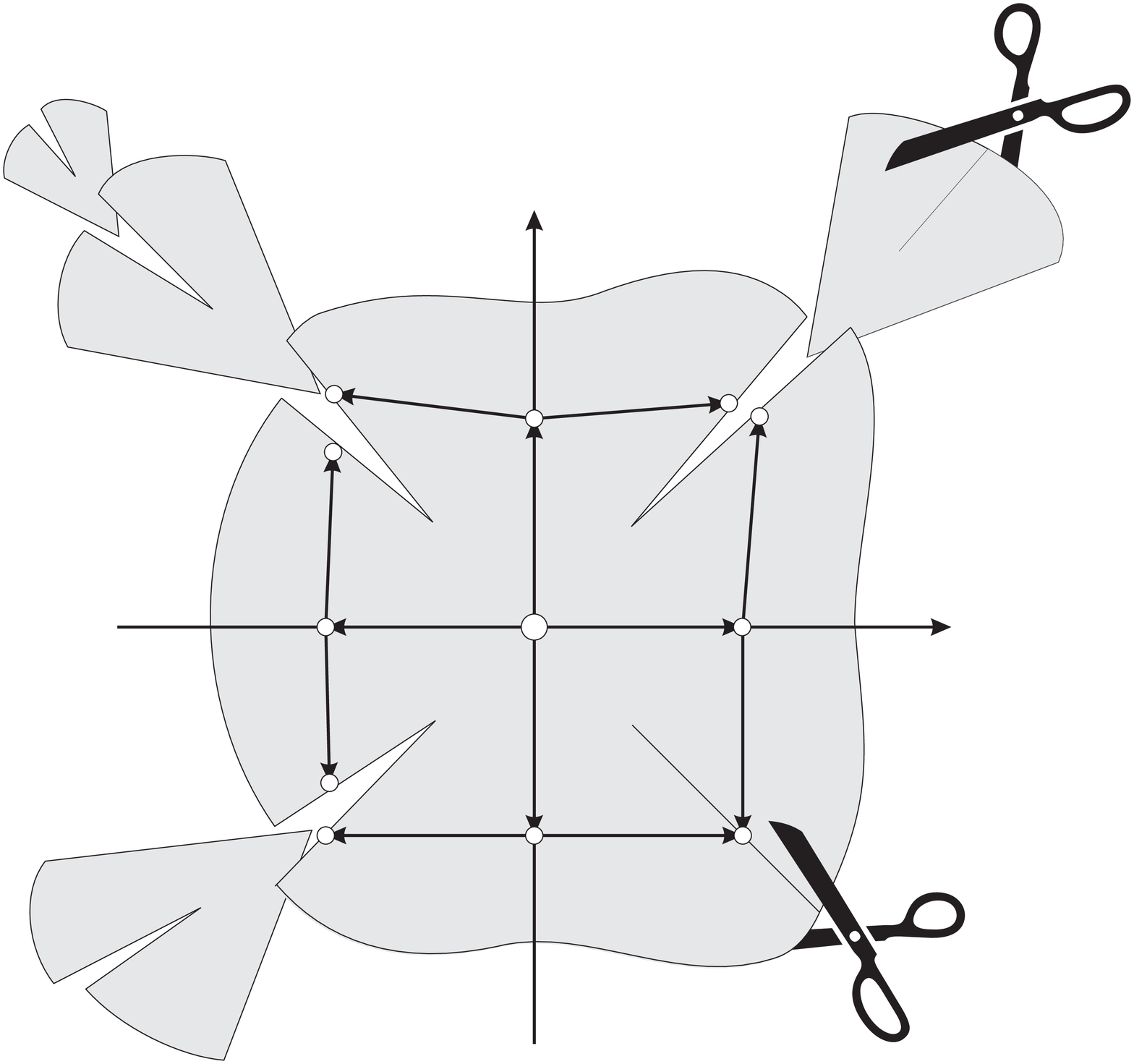,width=8cm}
\end{center}
\caption{(a) Picture of a burdock's leaf ({\it Arctium lappa}),
(b) The surface \`a godets isometrically covered by a 4--branching
Cayley tree.}
\label{fig:1}
\end{figure}

It is reasonable to imagine that cells located near the leaf's boundary
proliferate more active than ones in the bulk of the leaf. One possible
reason might be purely geometric: the periphery cells are not completely
enclosed by the surrounding media and hence have more available space for
growth than the cells inside the leaf's body\footnote{Despite this
explanation seems rather natural, some biological justifications of the
expressed hypothesis would be worthwhile.}. The corresponding picture of
the burdock leaf ({\it arctium lappa}) is reproduced in fig.\ref{fig:1}a.
This  phenomenon can be observed for various species such as most
of types of lettuce ({\it lactuca sativa}) or spinach ({\it spinacea
oleracea})... The  schematic model of the suggested geometric origin of the
leaf's folding is shown in fig.\ref{fig:1}b (see the explanations below).

At the same time the question (2) remained unanswered, and in this paper we
discuss some possible ways of solution to this problem. Relying on the
growth mechanism of a plant's leaf conjectured above, one can propose the
following naive construction schematically shown in fig.\ref{fig:1}b. Take
some ``elementary'' bounded domain of a flat surface, make finite radial
cuts and insert in these cuts flat triangles, modeling the area newly
generated by periphery cells. The resulting surface obtained by putting
back together all elementary domains and inserted extra triangles is not
flat anymore. Continue the process recursively, i.e. make cuts in the new
surface, insert extra flat triangles and so on... One  gets this way  a
surface whose perimeter and area grow exponentially with the radius.
Several examples can be found in \cite{god_int}.

Apparently a similar problem  had been discussed for the first time by the
Russian matematician P.L.Tchebychef in his talk "On the cut of clothes" in
Paris on 28 August 1878, apparently unpublished (some traces of this talk
we found in the web page \cite{w0}). Among the modern developments of that
subject one can mention the paper \cite{bowers} by P.Bowers proving in 1997
the theorem on quasi--isometric embedding of a uniform binary tree into a
negatively curved space, as well as the contribution \cite{duval} by
S.Duval and M.Tajine devoted to isometric embeddings of trees into metric
spaces for fractal description.

Our forthcoming consideration is based on two suppositions: (i) the plant's
leaf has infinitesimal thickness without any  surface tension, and (ii)
activity of boundary cells is independent on  the size of the leaf.

The surface constructed above is called "hyperbolic" \cite{0}. Anyone who
pays attention to the tendencies in fashion recognizes in this recursive
construction the so--called "jupe \`a godets". Later on we shall call such
surfaces the "surfaces \`a godets". Despite this rather transparent
geometrical image, the very problem under consideration is still too vague.
Let us formulate it in more rigorous terms, which allow the forthcoming
mathematical analysis.

Let us cover the surface \`a godets by a natural "lattice". The
construction of this "lattice" is as follows. Take a 4--branching Cayley
tree\footnote{The example of a 4--branching  Cayley tree is considered for
simplicity. In principle one can deal with any regular hyperbolic lattice.
Some other examples are discussed at length of Section \ref{sect:3}.}. It
is well known that any regular Cayley tree, as an exponentially growing
structure, cannot be isometrically embedded in any space with flat metric.
One can expect that the "surface \`a godets'', being by construction an
exponentially growing (i.e. ``hyperbolic'') structure, admits Cayley trees
as possible discretizations.  The reason for such choice is based on the
fact that the circumference $P(k)$ of the Cayley tree (i.e. the number of
outer vertices located at the distance $k$ from the tree root) grows as
$P(k)=z\times z^{k-1}$, where $z$ is the coordinational number of the
Cayley tree ($z=4$ for 4--branching tree). We shall assume that Cayley
trees can cover the "surface \`a godets" {\it isometrically}\footnote{For
example, the rectangular lattice shown in fig.\ref{fig:godets2}b
isometrically covers the Euclidean plane.}, i.e. without gaps and
selfintersections, preserving angles and distances---see figs.\ref{fig:1}b
and \ref{fig:godets2}a. Thus, our further aim consists in describing the
relief of the "surface \`a godets" in the 3--dimensional Euclidean space,
under the condition that a  regular 4--branching Cayley tree is
isometrically embedded in this surface.

\begin{figure}[ht]
\begin{center}
\epsfig{file=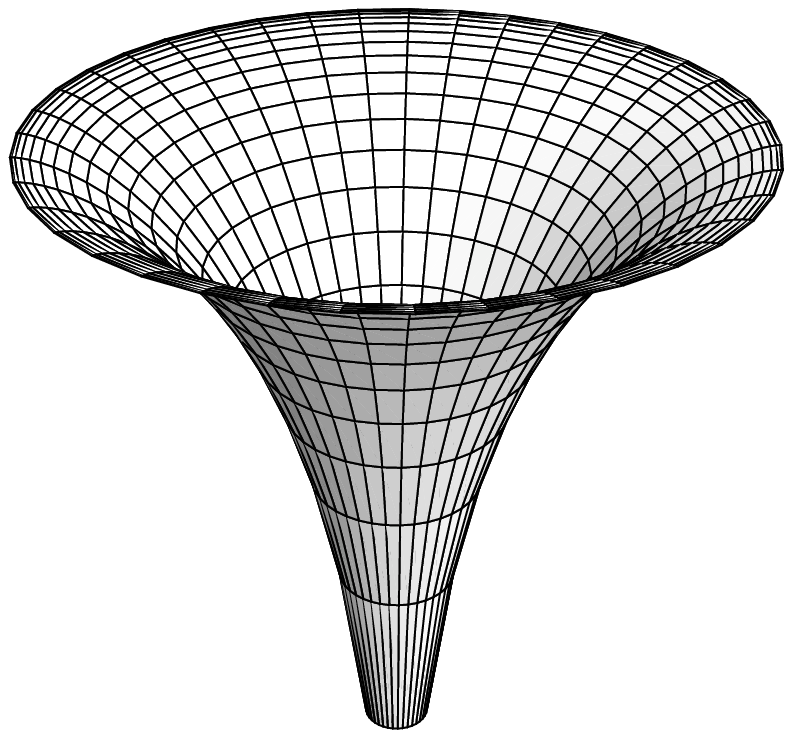,width=6cm}
\epsfig{file=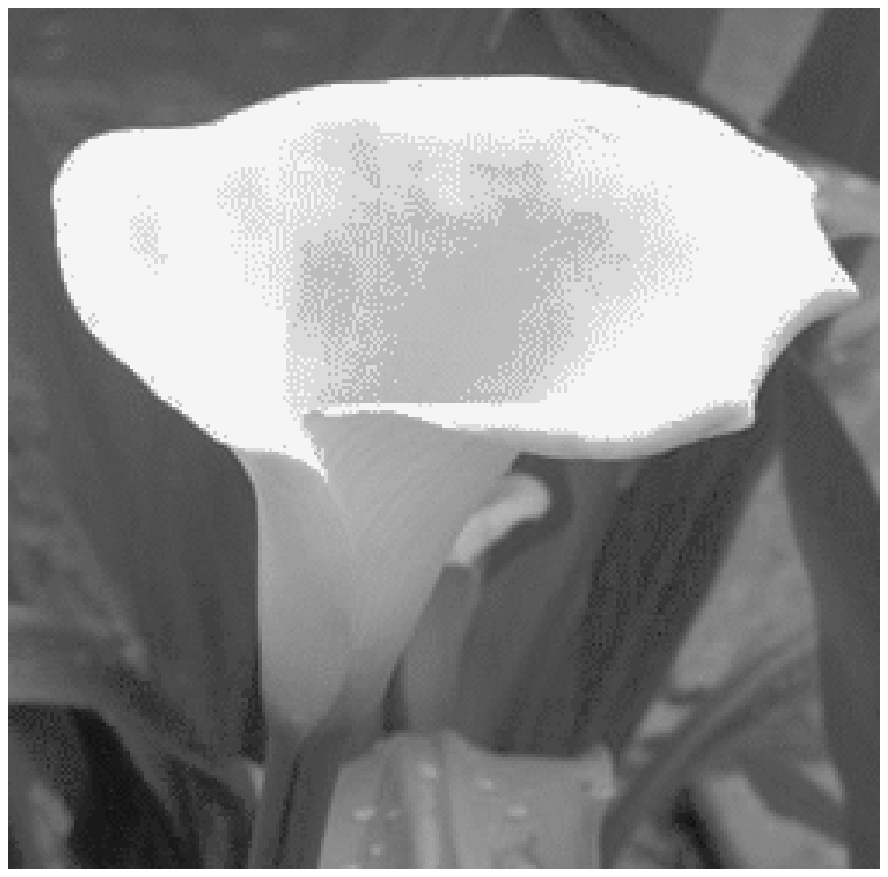,width=6cm}
\end{center}
\caption{(a) Pseudosphere, (b) Picture of calla lilly.}
\label{fig:pseudo}
\end{figure}

\noindent
{\it Comment.} \\
Let us stress that in our paper we are interested in embedding of open
(i.e. unbounded) surfaces only. From many textbooks on noneuclidean
geometry we know that exponentially growing surfaces {\it with the
boundary} can be isometrically embedded in a Euclidean space. The surface
of revolution of the so-called  {\it tractrix} around its asymptote is a
famous example of such embedding \cite{0}---see fig.\ref{fig:pseudo}a. In
3D Euclidean  space $x,y,z$ the pseudosphere is parametrized by the
following equations:
$$
\left\{
\begin{array}{l}
x=\cos u\, \sin v \\
y=\sin u\, \sin v \\
z=\cos v + \log\tan\frac{v}{2} \\
\end{array}
\right.
$$
where $0\le u\le 2\pi$ and $0<v\le\frac{\pi}{2}$.

One sees from fig.\ref{fig:pseudo}a that the surface of the pseudosphere
has no crumples, and as it is  bounded, it  contains a {\it finite part} of
a Cayley tree as an isometric embedding. We suspect that the pseudospheric
structure can be also realized in the plants world. The picture of the
flower of calla lilly reproduced in fig.\ref{fig:pseudo}b supports our
claim.

From our discussion, we conjecture that in the natural world both the
surface \`a godets (SG) and the pseudospere (PS) have equal rights to
exist. From the biological point of view there is no difference between SG
and PS---in both cases the same mechanism of cells proliferation is
assumed. However from the topological point of view the very crucial point is
the initial structure of the growing surface: if in the initial phase the
growing structure is"pipe--like",  then one can expect the pseudosphere
formation, while if the initial surface is almost flat and all
circumference cells are independently growing, we ultimately arrive at a
surface \`a godets.

After this comment we turn to the question of embedding isometrically a
4--branching {\it infinite} Cayley tree into a 2--manifold ("surface \`a
godets"), viewed as an open surface in the 3D Euclidean space.

\section{Isometric embedding of a "surface \`a godets" into the 3D Euclidean
space} \label{sect:2}

\subsection{The model}

Take a zero--angled rectangle $A_{\zeta}B_{\zeta}A'_{\zeta}C_{\zeta}$
bounded by arcs and lying in the unit disc $|\zeta|<1$ as shown in
fig.\ref{fig:conf}d. Make reflections of the interior of the rectangle with
respect to its sides $A_{\zeta}B_{\zeta}$, $B_{\zeta}A'_{\zeta}$,
$A'_{\zeta}C_{\zeta}$, $C_{\zeta}A_{\zeta}$ and get a new generation of
vertices -- the new images of the initial zero--angled rectangle \cite{1}.
For example, the reflection with respect to  the arc $A'_{\zeta}C_{\zeta}$ (which by definition remains unchanged) transforms $A_{\zeta}  \to A''_{\zeta}$ and $ B_{\zeta} \to  B'_{\zeta}$.

Proceeding recursively with reflections, one isometrically tessellates the
disc $|\zeta|<1$ with all images of the rectangle $A_{\zeta}B_{\zeta}
A'_{\zeta}C_{\zeta}$. If one connects the centers of neighboring (i.e.
obtained by successive reflections) rectangles, on gets a 4--branching
Cayley tree isometrically embedded into the unit disc endowed with the
Poincar\'e metric \cite{2} defined as follows:
$$
ds^2=\frac{d\zeta\,d\zeta^*} {(1-\zeta\,\zeta^*)^2}
$$
where $ds$ is the differential length and the variables $\zeta$ and
$\zeta^*$ are complex--conjugated.

It is known that the tessellation of the  Poincar\'e disc by circular
zero--angled rectangles is uniform in a surface of a 2--dimensional
hyperboloid obtained by stereographic projection from the unit disc
\cite{0}.  The open 2--hyperboloid is naturally embedded into a 3D space
with {\it Minkovski metric},  i.e. with metric tensor of signature
$\{+1,+1,-1\}$.  However  the problem of uniform embedding of an open
2--hyperboloid into a 3D space  with {\it Euclidean metric} deserves
special attention and exactly coincides with the main aim of our
work---embedding of a 4--branching  Cayley tree (isometrically covering the
unit disc with Poincar\'e metric)  into a 3D Euclidean space. Such
embedding is realized by a transform  $z=z(\zeta)$ which conformally maps a
flat square $A_zB_zA'_zC_z$ in the  complex plane $z$ to a circular
zero--angled rectangle $A_{\zeta}B_{\zeta} A'_{\zeta}C_{\zeta}$ in the unit
disc $|\zeta|<1$ (see  fig.\ref{fig:conf}). The relief of the corresponding
surface is encoded in  the so-called {\it coefficient of deformation}
$J(\zeta)= \left|\frac{dz}  {d\zeta}\right|^2$ coinciding with the Jacobian
of the conformal transform  $z(\zeta)$:
$$
J(\zeta)=|z'(\zeta)|^2=
\left|\begin{array}{cc}
\disp \frac{\partial\, {\rm Re}\, z}{\partial\, {\rm Re}\, \zeta} &
\disp \frac{\partial\, {\rm Re}\, z}{\partial\, {\rm Im}\, \zeta} \bigskip \\
\disp \frac{\partial\, {\rm Im}\, z}{\partial\, {\rm Re}\, \zeta} &
\disp \frac{\partial\, {\rm Im}\, z}{\partial\, {\rm Im}\, \zeta}
\end{array}\right|
$$
Our final goal consists in an explicit construction of the function
$J(\zeta)$.

\noindent
{\it Comments.} \\
(A) The Jacobian $J(\zeta)$ has singularities ($J(\zeta)= \infty$) at all
branching points---vertices $A_{\zeta},B_{\zeta}, A'_{\zeta}, C_{\zeta}$
and their images. \\
(B) One can easily show that all the images of zero--angled rectangle
$A_{\zeta}B_{\zeta}A'_{\zeta} C_{\zeta}$ in the complex plane $\zeta$ have
the same area. Namely, the area of the elementary cell (the rectangle 
$A_zB_zA'_zC_z$) tessellating the plane $z$ is
$$
S=\int\limits_{A_zB_zA'_zC_z}dz\,dz^* 
$$
Performing the transform $z(\zeta)$ one can rewrite $S$ as follows
$$
S=\int\limits_{A_{\zeta}B_{\zeta}A'_{\zeta} C_{\zeta}} J(\zeta)\,d\zeta\, 
d\zeta^*
$$
Taking into account the correspondence of boundaries under conformal 
transforms one arrives to the conclusion that all zero--angled rectangles 
have the same area $S$ in $\zeta$.

\subsection{Construction of conformal maps}

\begin{figure}[ht]
\begin{center}
\epsfig{file=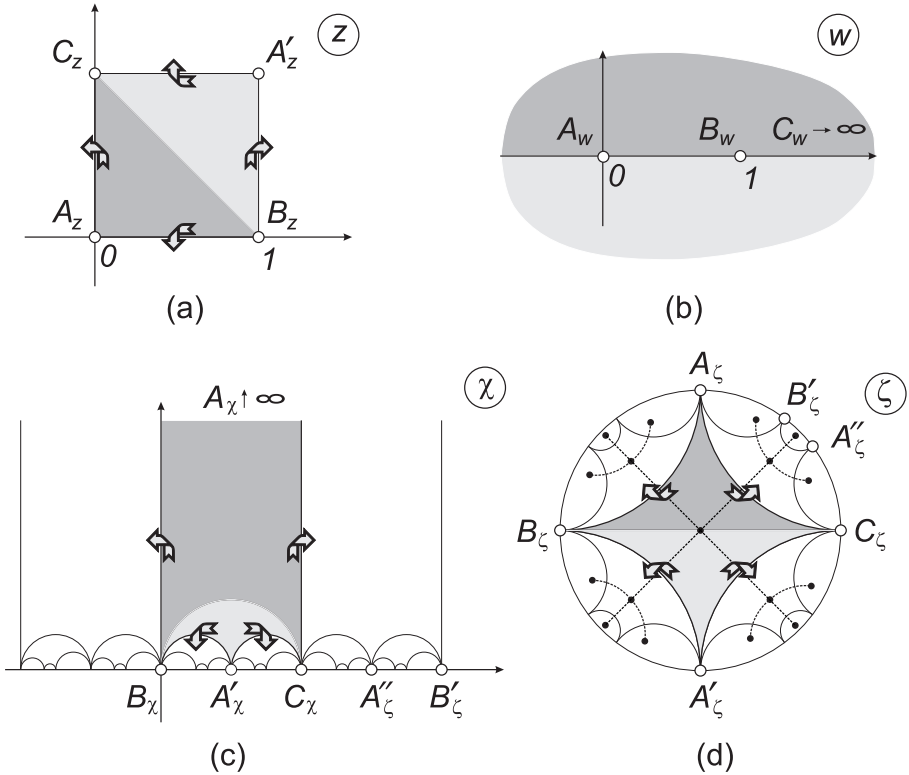, width=12cm}
\end{center}
\caption{Conformal maps: (a)--(b) $z=z(w)$; (b)--(c) $w=w(\chi)$;
(c)--(d) $\chi=\chi(\zeta)$; (d) The Cayley tree isometrically covering the
Poincar\'e disc $|\zeta|<1$ is shown by dotted line.}
\label{fig:conf}
\end{figure}

To find the conformal transform of the square $A_zB_zA'_zC_z$ in the $z$--plane
to the zero--angled square $A_{\zeta}B_{\zeta}A'_{\zeta}C_{\zeta}$ in the
$\zeta$--plane, consider the following sequence of auxiliary conformal maps
(shown in fig.\ref{fig:conf}):
$$
z \stackrel{z=z(w)}{\longrightarrow} w
\stackrel{w=w(\chi)}{\longrightarrow} \chi
\stackrel{\chi=\chi(\zeta)}{\longrightarrow}
\zeta
$$

Explicit construction of the functions $z(w)$, $w(\chi)$ and
$\chi(\zeta)$ is described below.

1. The function $z=z(w)$ conformally maps the interior of the triangle
$A_zB_zC_z$ with angles $\frac{\pi}{4},\frac{\pi}{4},\frac{\pi}{2}$ in the
plane $z$ onto the upper half--plane $w$ (see
fig.\ref{fig:conf}a). The conformal map $z=z(w)$ is performed via
Christoffel-Schwartz integral \cite{4}
\be \label{eq:conf1}
z(w)=\int_0^w
\frac{d\tilde{w}}{\tilde{w}^{1/2}\left(1-\tilde{w}\right)^{3/4}}
\ee
with the following correspondence of branching points (see
fig.\ref{fig:conf}a):
$$
\left\{\begin{array}{lcl}
A_z(z=0) & \to & A_w(w=0) \medskip \\
B_z(z=1) & \to & B_w(w=1) \medskip \\
C_z(z=i) & \to & C_w(w=\infty)
\end{array}\right.
$$

2. The function $w=w(\chi)$ conformally maps the upper half--plane ${\rm
Im}\,w>0$ to the interior of a circular triangle $A_{\chi}B_{\chi}
C_{\chi}$ with angles $(0,0,0)$ lying in the upper half--plane ${\rm
Im}\,\chi>0$ as it is shown in fig.\ref{fig:conf}b. The standard theory of
automorphic functions answers the question of explicitly constructing  the
map $w=w(\chi)$ (see, for example, \cite{5,6}).

Consider a function $u(w)$ satisfying the hypergeometric equation with three
branching points at $w=\{0,1,\infty\}$:
\be \label{eq:hyp}
w(w-1)u''(w)+\Big\{(\alpha+\beta+1)w-\gamma\Big\}u'(w)+
\alpha\beta\,u(w)=0
\ee
where the coefficients $\alpha,\,\beta,\,\gamma$ are uniquely defined by the
angles of the circular triangle $ABC$: $\{\mu_1\pi,\,\mu_2\pi,\,\mu_3\pi\}$:
$$
\mu_1=1-\gamma; \quad \mu_2=\gamma-\alpha-\beta; \quad \mu_3=\beta-\alpha
$$
In our case $\mu_1=\mu_2=\mu_3=0$, hence
$$
\alpha=\frac{1}{2}; \quad \beta=\frac{1}{2}; \quad \gamma=1
$$
The equation (\ref{eq:hyp}) has two linearly independent fundamental
solutions $u_1(w)$ and $u_2(w)$ which can be expressed in terms of
hypergeometric functions $F(\alpha,\beta,\gamma,z)$. For our particular 
choice of parameters $\alpha,\beta,\gamma$ eq.(\ref{eq:hyp}) belongs to 
so-called degenerate case, where:
\be \label{eq:hyp2}
\begin{array}{l}
u_1(w)=F\left(\frac{1}{2},\frac{1}{2},1,z \right) \medskip \\
u_2(w)=i F\left(\frac{1}{2},\frac{1}{2},1,1-z \right)
\end{array}
\ee

Recall that the function $F(\alpha,\beta,\gamma,z)$ admits the following 
integral representation
$$
F(\alpha,\beta,\gamma,z)=\frac{\Gamma(\gamma)}{\Gamma(\beta)
\Gamma(\gamma-\beta)}
\int_0^1 t^{\beta-1}(1-t)^{\gamma-\beta-1}(1-zt)^{-\alpha}dt
$$

It is known \cite{5,6,7} that the conformal map $\chi(w)$ of the interior 
of a circular triangle $A_{\chi}B_{\chi} C_{\chi}$ with angles $(0,0,0)$ 
lying in the upper half--plane ${\rm Im}\,\chi>0$ to the upper half--plane 
${\rm Im}\,w>0$ can be written as the quotient of fundamental solutions 
$u_1(w)$ and $u_2(w)$:
\be \label{eq:hyp3}
\chi(w)=\frac{u_1(w)}{u_2(w)}= -i \frac{F\left(\frac{1}{2},\frac{1}{2},1,z
\right)}{F\left(\frac{1}{2},\frac{1}{2},1,1-z \right)}
\ee
The conformal map of the circular triangle $A_{\chi}B_{\chi} C_{\chi}$ to
the upper half--plane ${\rm Im} w>0$ is mutually single--valued, thus the
inverse function $\chi^{-1}(w)$ solves our problem. Inverting the function
$\chi(w)$, (this inverse function satisfies a Schartzian equation, which can be solved in our case, as shown in \cite{7} p.23), we get:
\be \label{eq:hyp4}
w(\chi)=\frac{\vtheta_2^4(0,e^{i\pi \chi})}{\vtheta_3^4(0,e^{i\pi \chi})}
\ee
The conformal transform (\ref{eq:hyp4}) establishes the following
correspondence of branching points:
$$
\left\{\begin{array}{lcl}
A_w(w=0) & \to & A_{\chi}(\chi=\infty) \medskip \\
B_w(w=1) & \to & B_{\chi}(\chi=0) \medskip \\
C_w(w=\infty) & \to & C_{\chi}(\chi=1)
\end{array}\right.
$$

3. The function $\chi=\chi(\zeta)$ conformally maps the interior of
the circular triangle $A_{\chi}B_{\chi}C_{\chi}$ in the upper half--plane
${\rm Im}\,\chi>0$ to the interior of a circular triangle $A_{\zeta},
B_{\zeta},C_{\zeta}$ in the open unit disc $|\zeta|<1$ (see
fig.\ref{fig:conf}c). It is realized via the fractional transform
\be \label{eq:fract}
\chi(\zeta)=\frac{1-i}{2}\;\frac{\zeta+1}{\zeta-i}
\ee
with the following correspondence of branching points:
$$
\left\{\begin{array}{lcl}
A_{\chi}(\chi=\infty) & \to & A_{\zeta}(\zeta=i) \medskip \\
B_{\chi}(\chi=0) & \to & B_{\zeta}(\zeta=-1) \medskip \\
C_{\chi}(\chi=1) & \to & C_{\zeta}(\zeta=1)
\end{array}\right.
$$

Collecting eqs.(\ref{eq:conf1}), (\ref{eq:hyp4}) and (\ref{eq:fract}) we
arrive at a composite conformal map
$$
z(\zeta)=z\{w[\chi(\zeta)]\}
$$
The Jacobian $ J(\zeta)$ of the map $z(\zeta)$ reads
\be \label{eq:jac2}
\begin{array}{lll}
J(\zeta) & \equiv & \disp \left|\frac{d z(\zeta)}{d \zeta}\right|^2 =
\disp \left|\frac{d z(w)}{d w}\right|^2\, \left|\frac{d w(\chi)}{d
\chi}\right|^2\, \left|\frac{\chi(\zeta)}{d\zeta}\right|^2 \medskip \\
& = & \disp \frac{4}{\pi^2\,\left|\zeta-i\right|^4}\left|
\vtheta_1'\left(0,e^{i\pi \frac{1-i}{2} \frac{\zeta+1}{\zeta-i}}\right)
\right|^2\,\left|\vtheta_2\left(0,e^{i\pi \frac{1-i}{2} \frac{\zeta+1}
{\zeta-i}}\right)\right|^2
\end{array}
\ee
where (see \cite{8})
$$
\begin{array}{l}
\disp \vtheta_1(\tau,e^{i\pi\chi})=2e^{i\frac{\pi}{4}\chi}
\sum_{n=0}^{\infty} (-1)^n e^{i\pi n(n+1)\chi} \sin(2n+1)\tau \medskip \\
\disp \vtheta_1'(0,e^{i\pi\chi})\equiv \left.
\frac{\vtheta_1(\tau,e^{i\pi\chi})}{d\tau}
\right|_{\tau=0}=2e^{i\frac{\pi}{4}\chi} \sum_{n=0}^{\infty} (-1)^n (2n+1)
e^{i\pi n(n+1)\chi} \medskip \\
\disp \vtheta_2(0,e^{i\pi\chi})=2e^{i\frac{\pi}{4}\chi}
\sum_{n=0}^{\infty} e^{i\pi n(n+1)\chi}
\end{array}
$$

\subsection{Geometric structure of the leaf's boundary}

Let us parameterize the "coefficient of deformation" $J(\eta)\equiv
J(\eta,\varphi)$ by the variables $(\eta,\varphi)$, where $\eta$ and
$\varphi$ are correspondingly the hyperbolic distance and the
polar angle in the unit disc:
$$
\eta=\ln \frac{1+|\zeta|}{1-|\zeta|}; \quad
\varphi=\arctan\frac{{\rm Im}\,\zeta}{{\rm Re}\,\zeta}
$$

\begin{figure}[ht]
\begin{center}
\epsfig{file=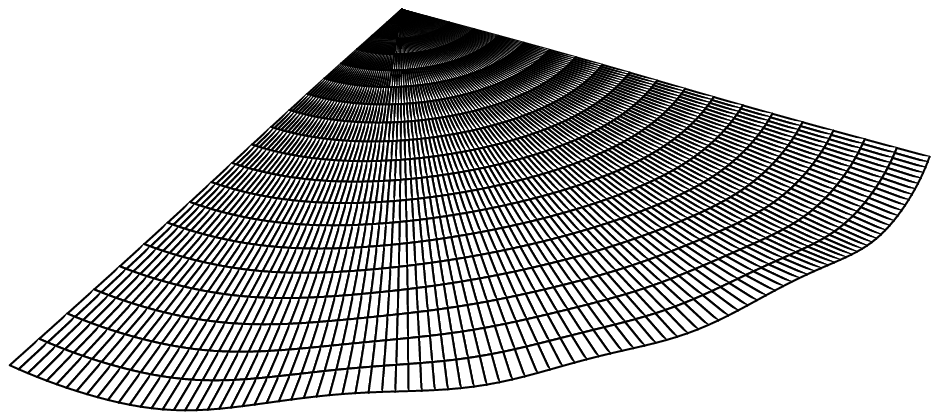,width=5cm} \hspace{-1.5cm}
\epsfig{file=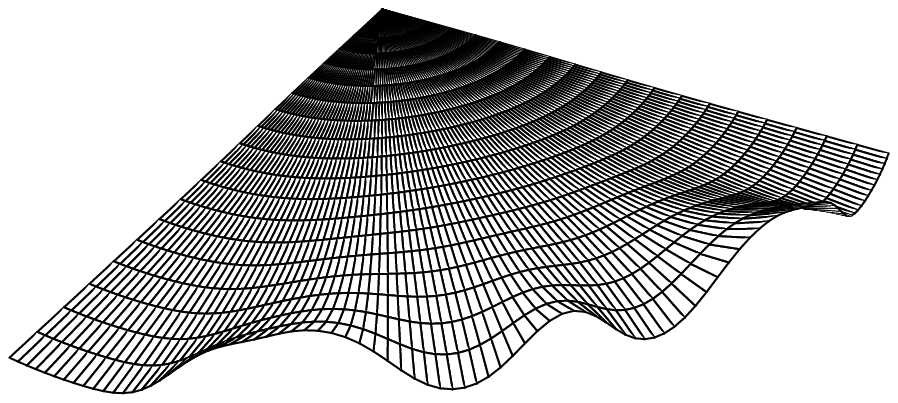,width=6cm} \hspace{-1.5cm}
\epsfig{file=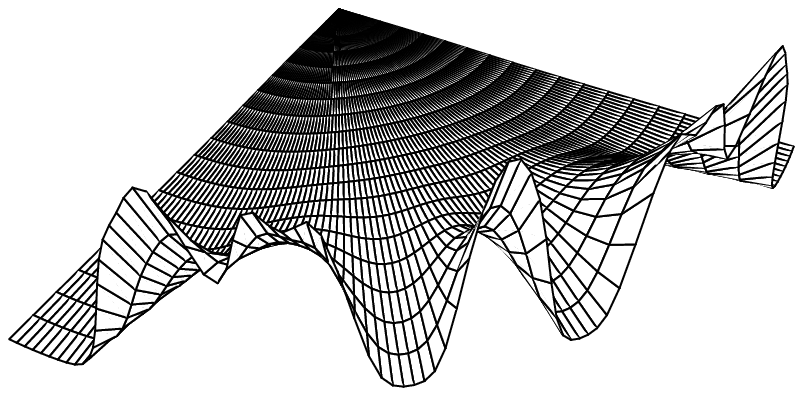,width=7cm}
\end{center}
\caption{Sample plots of $J(\eta,\varphi)$ for $0\le\eta\le \eta_{\rm max}$
and $0\le\varphi \le\frac{\pi}{2}$: (a) $\eta_{\rm max}=1.1$; (b)
$\eta_{\rm max}=1.4$; (c) $\eta_{\rm max}=1.7$.}
\label{fig:3}
\end{figure}

Few sample plots of the function $J(\eta,\varphi)$ for $0\le\eta\le
\eta_{\rm max}$ and $0\le\varphi \le\frac{\pi}{2}$ for $\eta_{\rm
max}=1.1;\,1.4;\,1.7$ are shown in fig.\ref{fig:3}a--c.

Equation (\ref{eq:jac2}) solves the problem of embedding a "jupe \`a godets"
isometrically covered by a 4--branching Cayley tree into the
three--dimensional Euclidean space. The corresponding boundary profile is 
shown in fig.\ref{fig:3}.

The growth of the perimeter $P(\eta)$ of the circular "surface \`a godets"
of hyperbolic radius $\eta$ as well as the span $h(\eta)$ of transversal
fluctuations can be characterized by the coefficients of growth ("Lyapunov
exponents"), $c_P$ and $c_h$ defined as follows
$$
c_P \equiv \lim_{\eta\to\infty}c_P(\eta)=
\lim_{\eta\to\infty}\frac{\ln P(\eta)}{\eta}; \quad
c_h \equiv \lim_{\eta\to\infty}c_h(\eta)=
\lim_{\eta\to\infty}\frac{\ln h(\eta)}{\eta}
$$
where
$$
P(\eta)=\int\limits_0^{2\pi}\sqrt{1+\left[\frac{dJ(\eta,\varphi)}
{d\varphi} \right]^2}d\,\varphi
$$
and
$$
h(\eta)=\int\limits_0^{2\pi}J(\eta,\varphi)\,d\varphi
$$

\begin{figure}[ht]
\begin{center}
\epsfig{file=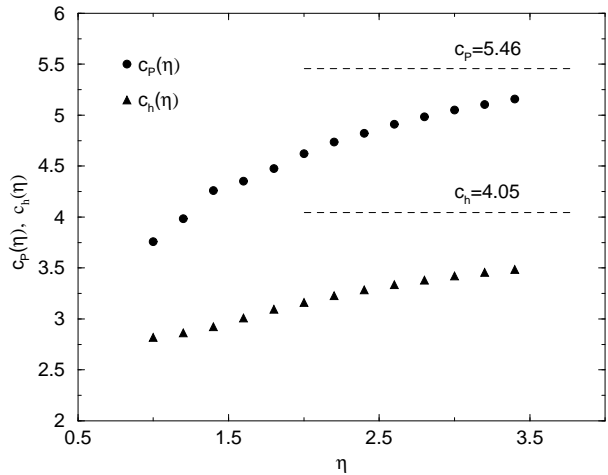, width=8cm}
\end{center}
\caption{Growth exponents $c_P(\eta)$ and $c_h(\eta)$.}
\label{fig:4}
\end{figure}

The plots of the functions $c_P(\eta)$ and $c_h(\eta)$ are shown in
fig.\ref{fig:4}.

\section{Growth of a random "surface \`a godets"}
\label{sect:3}

The natural generalization of the model of a regular "surface \`a godets" 
defined in Section \ref{sect:1} consists in its "randomization".  Suppose 
now that the growth phenomenon is random: extra area is generated only at 
randomly chosen (``active'') sites. It means that in our model we insert 
extra triangles with probability $1-q$ and leave an elementary domain 
unchanged with probability $q$. The vertices without inserted extra 
triangle we shall call "defects". They can be interpreted as defects of 
curvature, i.e. points where the surface is locally flat. It is easy to 
understand that if the probability of defects is  $q=0$, we return to the 
regular surface \`a godets with hyperbolic metric, while for $q=1$ the 
resulting surface is flat (i.e. it can be embedded in a plane). We are 
interested in the dependence of the perimeter $P(k|q)$ on the radial 
distance $k$ for the surface \`a godets with concentration of defects $q$.

Here again the problem becomes much more transparent if formulated for a 
discrete realization (isometrically embedded graph) of the ``surface \`a 
godets''. The issue is to define defects of curvature for graphs. First of 
all let us recall the definition of the Cayley graph of a group $G$.

By definition, the graph of the group $G$ is constructed as follows:
\begin{itemize}
\item The vertices of the graph are labelled by group elements. Every 
element of the group is represented by some irreducible (i.e. written with 
the minimal number of letters) word (not necessarily unique) built from the 
letters -- generators of the group $G$. Two  words equivalent in the group $G$ correspond to one and the same vertex of the graph.
\item Two different vertices corresponding to nonequivalent words $w$ and 
$w'$ are connected by an edge (i.e. are nearest--neighbors) if and only if 
the word $w'$ can be obtained from the word $w$ by deleting one letter 
(generator of the group $G$), or adding one extra letter. 
\end{itemize}

The following construction is of use.

\begin{figure}
\begin{center}
\epsfig{file=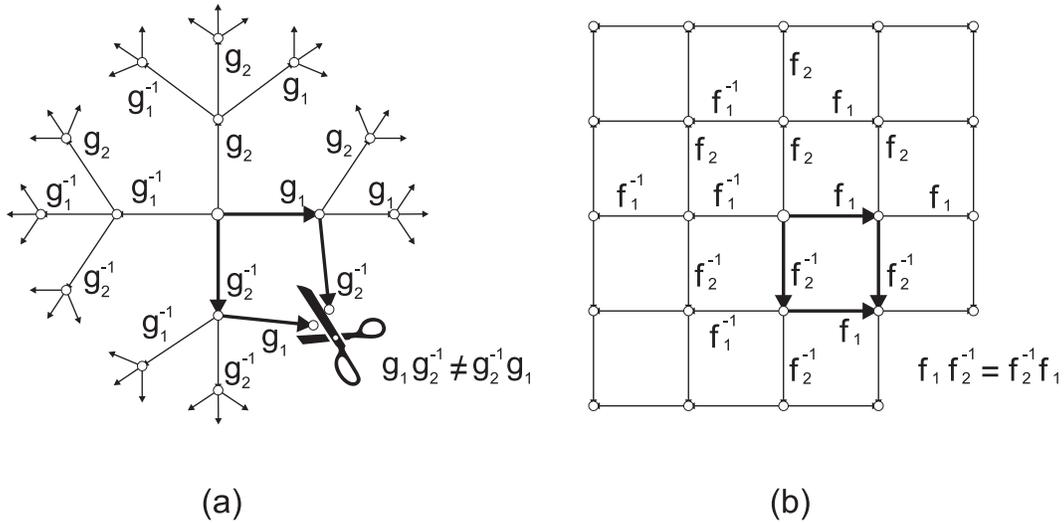,width=14cm}
\end{center}
\caption{(a) The 4--branching Cayley graph; (b) The graph (lattice)
covering a surface with Euclidean (flat) metric.}
\label{fig:godets2}
\end{figure}

1. It is known that a 4--branching Cayley tree is the graph of the {\it free}
group $\Gamma_2$ (see fig.\ref{fig:godets2}a). The group $\Gamma_2$ is the
infinite group of all possible words constructed from the set of letters
$\{g_1,g_2,\,g_1^{-1},g_2^{-1}\}$, where there are {\it no} commutation
relations among the letters \cite{1,9}. The total number of nonequivalent
shortest (irreducible) words of length $k$ in the group $\Gamma_2$ equals to
the number $P_{\Gamma_2}(k)$ of distinct vertices of the 4--branching Cayley
tree lying at a distance $k$ from the origin (see fig.\ref{fig:godets2}a):
\be \label{2:vfree}
P_{\Gamma_2}(k)=4\times 3^{k-1} \qquad (k\ge 1)
\ee

2. Consider now the opposite case of the {\it completely commutative 
group} $E_2$. This group is generated by $\{f_1,f_2,\,f_1^{-1},f_2^{-1}\}$, 
with the relation $f_1f_2=f_2f_1$. The graph of the group $E_2$ is a 
4--vertex lattice isometrically covering the Euclidean plane (see 
fig.\ref{fig:godets2}b). The problem of comparing two words $w_1$ and $w_2$ 
in the group $E_2$ written in different ways has a very straightforward 
solution. Taking into account that all generators of the group $E_2$ 
commute, we may write all irreducible words  of length $k$ in an {\it 
ordered form}:
$$
w=g_1^{m_1}\,g_2^{m_2}\qquad (|m_1|+|m_2|=k)
$$
where
$$
\begin{array}{l}
m_1=\#[\mbox{number of generators $f_1$}]-\#[\mbox{number of generators
$f_1^{-1}$}] \\
m_2=\#[\mbox{number of generators $f_2$}]-\#[\mbox{number of generators
$f_2^{-1}$}]
\end{array}
$$

It is convenient to encode the growth of the number of irreducible words 
with their length by an ``incident  matrix'' $\hat{T}$. This construction 
relies on the automatic structure (see \cite{charney}) of the groups under 
consideration.  All words of length $k+1$ are obtained by right hand  
multiplication of words of length $k$ by admissible generators. The 
admissible generators for a word $w$ depend {\it only} on the last letter 
of $w$. In our case the entry $(i,j)$ of the matrix $\hat{T}$ is 1 if $f_j$ 
is admissible after $f_i$ and 0 otherwise. Let us note that the 
corresponding construction has been used recently in \cite{10} in the case 
of "locally free groups". For the commutative group $E_2$ the incident 
matrix reads:
\be \label{eq:tra1}
\hat{T}=
\begin{tabular}{|c|c|c|c|c|} \hline
& $f_1$ & $f_2$ & $f_1^{-1}$ & $f_2^{-1}$ \\ \hline
$f_1$ & 1 & 1 & 0 & 1 \\ \hline
$f_2$ & 0 & 1 & 0 & 0 \\ \hline
$f_1^{-1}$ & 0 & 1 & 1 & 1 \\ \hline
$f_2^{-1}$ & 0 & 0 & 0 & 1 \\ \hline
\end{tabular}
\ee
The total number of shortest words of length $k$ is
\be \label{eq:comm}
P(k)= {\bf v} \left[\hat{T}\right]^{k-1} {\bf v}^{\top}= 4k \qquad (k\ge 1)
\ee
where ${\bf v}=(1,1,1,1)$ and ${\bf v}^{\top}$ is the transposed vector.

Our idea to mimic the ``surface \`a godets'' with defects of curvature 
consists in the following. Let us insert ``defects'' in the commutation 
relations of the  free group $\Gamma_2$. It means that passing ``along a 
word'', each time one meets a pair of consecutive generators with different 
subscripts (2 and 1), one either commutes them with probability $q$, or 
leave the sequence without changes with probability $1-q$. The similar idea 
has been developed in \cite{11} to approximate braid groups. More 
explicitely, one makes the  following substitutions:
\be \label{eq:defect}
\begin{array}{cc}
g_2\, g_1 \to
\left\{\begin{array}{ll}
g_1\, g_2 & \mbox{with prob. $q$} \\
g_2\, g_1 & \mbox{with prob. $1-q$}
\end{array}\right.; &
g_2^{-1}\, g_1 \to
\left\{\begin{array}{ll}
g_1\, g_2^{-1} & \mbox{with prob. $q$} \\
g_2^{-1}\, g_1& \mbox{with prob. $1-q$}
\end{array}\right. \bigskip \\
g_2\, g_1^{-1} \to
\left\{\begin{array}{ll}
g_1^{-1}\, g_2 & \mbox{with prob. $q$} \\
g_2\, g_1^{-1} & \mbox{with prob. $1-q$}
\end{array}\right.; &
g_2^{-1}\, g_1^{-1} \to
\left\{\begin{array}{ll}
g_1^{-1}\, g_2^{-1} & \mbox{with prob. $q$} \\
g_2^{-1}\, g_1^{-1} & \mbox{with prob. $1-q$}
\end{array}\right.
\end{array}
\ee

For $q=0$ we recover the commutation relations of  the free group $\Gamma_2$
with exponentially growing "volume" (the number of nonequivalent words)
$P_{\Gamma_2}(k|q=1)$, while for $q=1$ we arrive at the  group $E_2$ whose  
"volume" $P_{E_2}(k|q=1)$ displays  a polynomial growth in $k$. Thus,
$$
\begin{array}{l}
\disp v_{\Gamma_2}\equiv v(q=0)=\lim_{k\to\infty}\frac{\ln P_{\Gamma_2}
(k|q=0)}{k}=
\ln 3 > 0 \medskip \\
\disp v_{E_2}\equiv v(q=1)=\lim_{k\to\infty}\frac{\ln P_{E_2}(k|q=1)}{k}=0
\end{array}
$$

Let us insert "defects" in commutation relations of the incident matrix
$\hat{T}(x_1,...,x_4)$:
\be \label{eq:tra}
\hat{T}(x_1,...,x_4)=
\begin{tabular}{|c|c|c|c|c|} \hline
& $g_1$ & $g_2$ & $g_1^{-1}$ & $g_2^{-1}$ \\ \hline
$g_1$ & 1 & 1 & 0 & 1 \\ \hline
$g_2$ & $x_1$ & 1 & $x_2$ & 0 \\ \hline
$g_1^{-1}$ & 0 & 1 & 1 & 1 \\ \hline
$g_2^{-1}$ & $x_3$ & 0 & $x_4$ & 1 \\ \hline
\end{tabular}
\ee
where
\be \label{eq:tra2}
x_m=\left\{\begin{array}{ll}
0 & \mbox{with probability $q$} \\
1 & \mbox{with probability $1-q$}
\end{array}\right.
\ee
and $x_m$ are independent for all $1\le m\le 4$.

The function $P(k|x_1^{(1)},...,x_4^{(1)},...,x_1^{(k)},...,x_4^{(k)})$
describing the volume growth in an ensemble of words with random
commutation relations, reads now (compare to (\ref{eq:comm}))
\be \label{eq:wq}
P(k|x_1^{(1)},...,x_4^{(1)},...,x_1^{(k)},...,x_4^{(k)})=
{\bf v}\left[\prod_{i=1}^{k-1}
\hat{T}(x_1^{(i)},...,x_4^{(i)})\right]{\bf v}^{\top}
\ee

We shall distinguish between two distribution of defects: (a) annealed and
(b) quenched. In case (a) the partition function $P(k|x_1^{(1)},...,
x_4^{(1)},...,x_1^{(k)},...,x_4^{(k)})$ is averaged with the measure
(\ref{eq:tra2}), while in case (b) the "free energy", i.e. the logarithmic
volume $v\sim \ln P$ is averaged with the same measure.

In the rest of the paper we pay attention to the case (a) only. Averaging
the function $P(k|x_1^{(1)},..., x_4^{(1)},...,x_1^{(k)},...,x_4^{(k)})$
over the disorder we get
\be
P(k|q)\equiv\left<P(k|x_1^{(1)},..., x_4^{(1)},...,x_1^{(k)},...,
x_4^{(k)})\right>={\bf v} \left<\hat{T}(x_1,..., x_4)\right>^{k-1}
{\bf v}^{\top}
\ee
where
\be
\begin{array}{lll}
\left<\hat{T}(x_1,...,x_4)\right> & = & \disp \sum_{x_1=0}^{1}...
\sum_{x_4=0}^{1} q^{4-(x_1+x_2+x_3+x_4)}(1-q)^{x_1+x_2+x_3+x_4}
\hat{T}(x_1,x_2,x_3,x_4)
\medskip \\ & = & \left(\begin{array}{cccc}
1 & 1 & 0 & 1 \\
1-q & 1 & 1-q & 0 \\
0 & 1 & 1 & 1 \\
1-q & 0 & 1-q & 1
\end{array}\right)
\end{array}
\ee
The highest eigenvalue $\lambda_{\rm max}$ of the averaged matrix
$\left<\hat{T}(x_1,...,x_4)\right>$ defines the logarithmic volume
$v^{a}(q)=\lim\limits_{k\to\infty}\frac{\ln P(k|q)}{k}$ of the system with
"annealed" defects in commutation relations:
\be
v^{a}(q)=\ln\left(1+2\sqrt{1-q}\right)
\ee
One can check that for any $0\le q<1$ the system is caracterized by an 
exponential growth, i.e. $v^a(q)>0$. However at $q=q_{\rm cr}=1$ one has a 
transition to a flat metric with $v^a(q=1)=0$. Expanding $v^a(q)$ in the 
vicinity of the transition point $q_{\rm cr}=1$, one gets
\be \label{eq:vga}
v^a(q)\Big|_{q\to 1^-}=2\,(1-q)^{1/2}+O\left[(1-q)^{3/2}\right]
\ee
In the framework of the standard classification scheme, the phase 
transitions are distinguished by the scaling exponent $\alpha$ in 
the behavior of the free energy $F(\tau)$ near the critical point 
$\tau_{cr}$ \cite{12}: 
\be \label{eq:f} 
F(\tau)\Big|_{\tau\to\tau_{\rm cr}}\propto |\tau_{\rm cr}-\tau|^{\alpha} 
\ee
Comparing (\ref{eq:f}) and (\ref{eq:vga}) one can attribute (\ref{eq:vga}) 
to a phase transition with the critical exponent $\alpha=\frac{1}{2}$.

\section{Discussion}
\label{sect:4}

\begin{figure}[ht]
\begin{center}
\epsfig{file=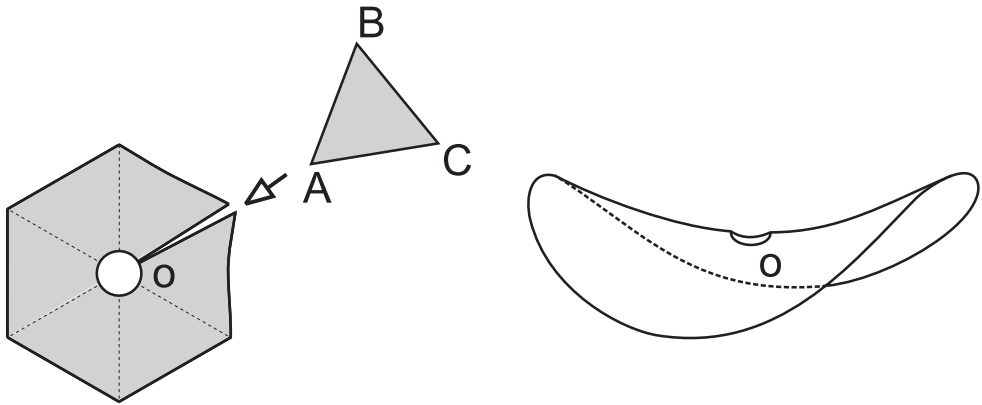, width=13cm}
\end{center}
\caption{A hexagon with a radial cut and inserted extra triangle and a
saddle--like structure near the corner point.}
\label{fig:5}
\end{figure}

In Sections \ref{sect:1} and \ref{sect:2} we have developed a method based 
on conformal transforms allowing to describe the growth of a regular ``jupe 
\`a godets''. Let us stress that our derivation is based on the assumption 
that the ``jupe \`a godets'' is isometrically covered by a 4--branching 
Cayley tree. We do not pretend to cover with this construction all possible 
surfaces \`a  godets (let us recall that an exponentially growing structure 
like a ``jupe \`a godets'' has infinitely many discrete lattices of 
isometries), but to provide an analytical tool for a quantitative 
description of the main features of such surfaces. It would be in a sense 
more natural to suggest a construction, corresponding to another 
hyperlattice (so--called $\{3,7\}$--lattice \cite{gut}). Take the right 
hexagon consisting of 6 equal--sided triangles and make a radial cut along 
a side of one triangle as it is shown in fig.\ref{fig:5}. In a given cut 
insert an extra triangle such that the central point is surrounded now by 7 
triangles; the resulting structure can be tentatively denoted as the 
``right 7--gon''. It is clear that such structure cannot be embedded in a 
plane anymore and has locally (near the center of a 7--gon) a saddle--like 
structure---see fig.\ref{fig:5}.

Let us continue the surface construction and glue new equal--sided 
triangles to all perimeter bonds of a given 7--gon in such a way that each 
corner of any triangle in the surface is surrounded exactly by 7 triangles. 
Hence near each corner point the surface has a saddle--like structure. In 
what follows we shall call this structure the ``7--gon surface \`a 
godets''.

\begin{figure}[ht]
\begin{center}
\epsfig{file=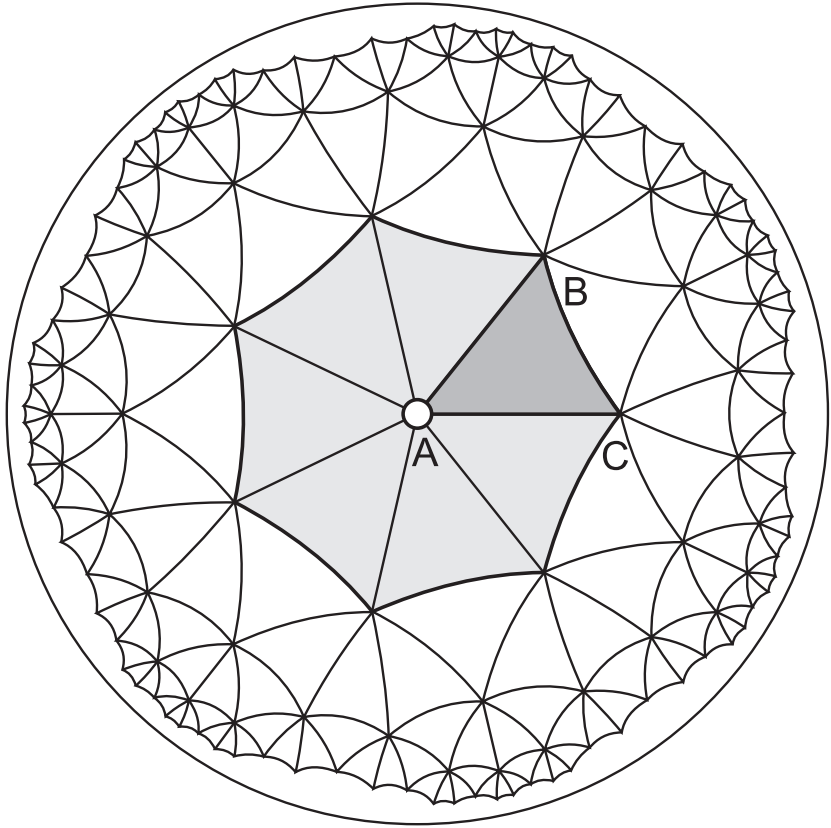, width=11cm}
\end{center}
\caption{A tessellation of a Poincar\'e disc by triangles with angles
$\left\{\frac{2\pi}{7},\,\frac{2\pi}{7},\,\frac{2\pi}{7}\right\}$.}
\label{fig:6}
\end{figure}

Consider now the unit disc endowed with the hyperbolic Poincar\'e metric 
and take the equal--sided circular (i.e. bounded by geodesics) triangle 
$ABC$ with angles $\left\{\frac{2\pi}{7},\,\frac{2\pi}{7},\,\frac{2\pi}{7}
\right\}$ as shown in fig.\ref{fig:6}.

Make reflections of the interior of the triangle $ABC$ with respect to its
sides and get the images of the initial triangle, then make reflections of
images with respect to their own sides and so on... In such a way one
tessellates the whole disc by the images of the initial triangle $ABC$. It is
easy to establish a topological bijection between the model of glued 7--gons
and the system of hyperbolic triangles with angles $\left\{\frac{2\pi}{7},\,
\frac{2\pi}{7},\, \frac{2\pi}{7}\right\}$, because all vertices are surrounded
exactly by 7 curvilinear hyperbolic triangles---see fig.\ref{fig:6}.

The structure shown in fig.\ref{fig:6} is invariant under conformal 
transforms of the Poincar\'e disc onto itself 
$$ 
\zeta\to \frac{\zeta -\zeta_0}{\zeta\zeta^*_0-1} 
$$ 
where $\zeta$ and $\zeta^*$ are complex--conjugated and $\zeta^*_0$ is the 
coordinate of any image of the point $0$ (vertex $A$ of triangle $ABC$). 
Such conformal transform allows to shift any image of a vertex of any
triangle to the center $0$ of the  disc (hence, such conformal transform 
replaces translations in Euclidean geometry).

It would be desirable to develop the symbolic language allowing to 
investigate statistical properties of a ``surface \`a godets'' with and 
without random ``defects of curvature'' for the structure shown in 
fig.\ref{fig:6}. We expect that the symbolic language similar to the one 
developed in Section \ref{sect:3} applied to the group of reflections of 
triangles with angles $\left\{\frac{2\pi}{7},\, \frac{2\pi}{7},\, 
\frac{2\pi}{7}\right\}$ would allow such generalization.

Let us conclude with the following remark. The conformal embedding of a 
``7-gon surface \`a godets'' shown in fig.\ref{fig:6} into a 3D Euclidean 
space can be algorithmically solved in the same way as it has been done for 
the 4--branching Cayley tree. However for the ``7--gon surface'' we cannot 
derive an explicit expression for the coefficient of deformation $J(\zeta)$ 
in a simple form (like in eq.(\ref{eq:jac2})). We outline this construction 
following results of Section \ref{sect:2}. Let $\tilde{z}(\tilde{\zeta})$ 
be the conformal transform of the triangle $\tilde{A}_z \tilde{B}_z 
\tilde{C}_z$ with angles $\left\{\frac{\pi}{3},\, \frac{\pi}{3},\, 
\frac{\pi}{3}\right\}$ in the complex plane $z$ to the triangle 
$\tilde{A}_{\zeta} \tilde{B}_{\zeta} \tilde{C}_{\zeta}$ with angles 
$\left\{\frac{2\pi}{7},\,\frac{2\pi}{7},\, \frac{2\pi}{7}\right\}$ in the 
Poincar\'e disc $|\zeta|<1$. The Jacobian $J(\tilde{\zeta})$ solves the 
problem of isometric embedding of a ``7--gon surface \`a godets'' into the  
3D Euclidean space. For the proper choice of parameters $\alpha,\beta,
\gamma$ in (\ref{eq:hyp}) we can derive the fundamental solutions $u_1(w)$ 
and $u_2(w)$ (and, hence $\chi(w)$), however unfortunately we are unable to 
find the inverse function $w(\chi)$. An explicit expression for 
$J(\tilde{\zeta})$ is therefore out of reach so far.

\bigskip

\centerline{\bf Acknowledgments}
\medskip

The authors are grateful to Vladimir Zakharov for drawing their attention 
to the problem as well as to Alexander Grosberg and Nicolas Rivier for 
stimulating discussions.

\end{document}